\documentclass[twocolumn,secnumarabic,amssymb,nobibnotes,aps,pre,superscriptaddress]{revtex4-1}

\usepackage{graphicx}
\usepackage{amsmath}

\newcommand{\nn}{\nonumber}
\newcommand{\eq}[1]{Eq.~\eqref{#1}}
\newcommand{\eqs}[1]{Eqs.~\eqref{#1}}
\newcommand{\fig}[1]{Fig.~\ref{#1}}

\newcommand{\eqsi}[1]{Eq.~SI\eqref{#1}}
\newcommand{\eqssi}[1]{Eqs.~SI\eqref{#1}}
\newcommand{\figsi}[1]{Fig.~SI\ref{#1}}
\newcommand{\appndx}[1]{SI~\ref{#1}}
\newcommand{\movie}[1]{Movie~\ref{#1}}

\newcommand{\ie}{\textit{i.e.} }

\newcommand{\tens}[1]{\boldsymbol{\mathbf{#1}}}
\newcommand{\del}{\vec{\nabla}}
\newcommand{\abs}[1]{\left| #1 \right|}
\newcommand{\av}[1]{\left\langle #1 \right\rangle}

\newcounter{siequation}
\newcounter{sifigure}
\newcounter{sisection}
\newcounter{simovie}
\usepackage[dvipsnames]{xcolor}

\usepackage{tikz}
  \newlength{\squareheight}
  \newcommand{\squareslash}{\tikz{\draw (0,0) rectangle (\squareheight,\squareheight);\draw(0,0) -- (\squareheight,\squareheight)}}
  \DeclareMathOperator{\squarediv}{\squareslash}

\begin{document}
\title{Helical Flow States in Active Nematics}

\author{Ryan Keogh}
\affiliation{School of Physics and Astronomy, The University of Edinburgh, Peter Guthrie Tait Road, Edinburgh, EH9 3FD, UK.}
\author{Santhan Chandragiri}
\affiliation{Department of Chemical Engineering, Indian Institute of Technology Madras, Chennai 600036, India.}
\author{Benjamin Loewe}
\affiliation{School of Physics and Astronomy, The University of Edinburgh, Peter Guthrie Tait Road, Edinburgh, EH9 3FD, UK.}
\author{Tapio Ala-Nissila}
\affiliation{MSP Group, QTF Centre of Excellence, Department of Applied Physics, Aalto University, P.O. Box 11000, FI-00076 Aalto, Espoo, Finland}
\affiliation{Interdisciplinary Centre for Mathematical Modelling, Department of Mathematical Sciences, Loughborough University, Loughborough LE11 3TU, United Kingdom}
\author{Sumesh Thampi}
\affiliation{Department of Chemical Engineering, Indian Institute of Technology Madras, Chennai 600036, India.}
\author{Tyler N. Shendruk}
\email{t.shendruk@ed.ac.uk}
\affiliation{School of Physics and Astronomy, The University of Edinburgh, Peter Guthrie Tait Road, Edinburgh, EH9 3FD, UK.}

\begin{abstract}
We show that confining extensile nematics in 3D channels leads to the emergence of two self-organized flow states with nonzero helicity. 
The first is a pair of braided anti-parallel streams---this {\em double helix} occurs when the activity is moderate, anchoring negligible and reduced temperature high. 
The second consists of axially aligned counter-rotating vortices---this {\em grinder train} arises between spontaneous axial streaming and the vortex lattice. 
These two unanticipated helical flow states illustrate the potential of active fluids to break symmetries and form complex but organized spatio-temporal structures in 3D fluidic devices.
\end{abstract}

\maketitle

\begin{figure}[b]
	\centering
	\includegraphics[width=0.5\textwidth]{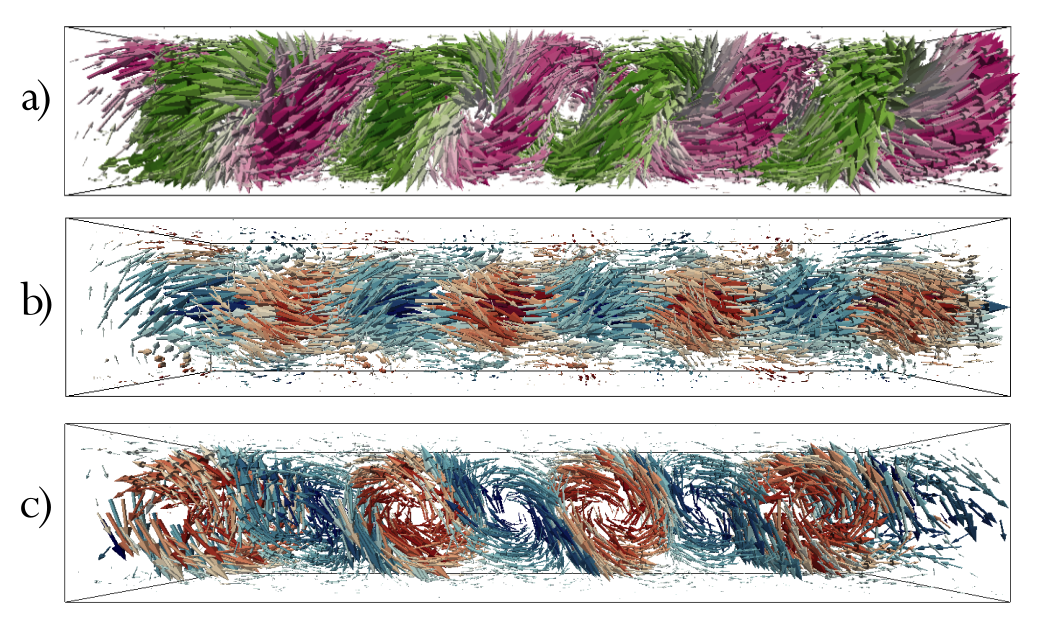}
	\caption{\small
	\textbf{Active dissipative flow structures exhibiting rotational flow.}
 	\textbf{(a) Double helix:} Two streams of anti-parallel flow braid around one another for parameters $\tilde{P} = \{ \tilde{K}^{-1/2},\tilde{W},\tilde{T}^{1/2} \}=\{ 25,1.5,9.47\}$. Coloured by axial velocity $u_\parallel$ (\movie{mov:DH}).
 	\textbf{(b) Grinder train:} Procession of spontaneously drifting, counter-rotating vortices aligned axially down the duct for $\tilde{P} = \{ 20,250,0 \}$. Coloured by axial vorticity $\omega_\parallel$ (\movie{mov:GT}).
 	\textbf{(c) Ceilidh vortex:} Vortex lattice oriented in the transverse direction for $\tilde{P}=\{ 22,250,0 \}$. Coloured by transverse vorticity $\omega_\perp$ (\movie{mov:VL}). 
	}
	\label{fig:flows}
\end{figure}


Because active materials are composed of microscopic constituents, which locally transmute internal energy into mechanical forces that are dissipated by the surrounding medium, they can possess broken symmetries and emergent dynamic properties on macroscopic scales. 
These self-organized, {\em active dissipative structures} have been observed to take many forms. 
The premier example of activity-induced dissipative structures is flocking of polar particles~\cite{vicsek1995,TonerTu1995,TonerTu1998}, in which a liquid-gas-like transition~\cite{solon2015} from disorderly to collective motion is associated with spontaneous spatial phase separation~\cite{chate2008}. 
In addition to coexistence, hexatic, smectic and solid phases can originate from activity~\cite{Digregorio2018,Chattopadhyay2021}. 
Likewise, activity can generate orientational order~\cite{santhosh2020,Giordano2021}, including active alignment of motile rods~\cite{bar2020,nagel2020} and swimming bacilliforms~\cite{nishiguchi2017}. 

\begin{figure*}[tb]
	\centering
	\includegraphics[width=0.99\textwidth]{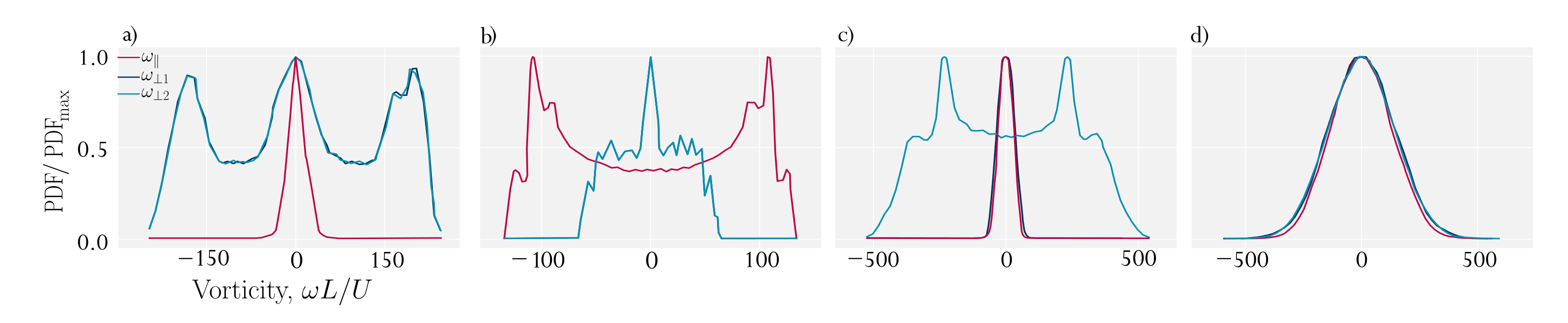}
	\caption{\small
	\textbf{Probability distributions of vorticity for active dissipative flow structures at parameters $\tilde{P} = \{ \tilde{K}^{-1/2},\tilde{W},\tilde{T}^{1/2} \}$.} 
	The components are the vorticity taken about the center line in the axial direction $\omega_\parallel$ (red lines) and in the two transverse directions $\omega_{\perp_{1,2}}$ (blue and cyan lines) averaged over ten random initializations. 
 	\textbf{(a) Double helix} ($\tilde{P} = \{ 25,1.5,9.47 \}$) possesses non-zero vorticity in both transverse directions $\omega_{\perp_{1,2}}$.
 	\textbf{(b) Grinder train} ($\tilde{P} = \{ 20,250,0 \}$). 
 	\textbf{(c) Ceilidh vortex lattice} ($\tilde{P} = \{ 21,250,0 \}$), possesses non-zero vorticity only in one of the two transverse directions. 
 	\textbf{(d) Turbulent flow} ($\tilde{P} = \{ 25,250,0 \}$) is nearly isotropic.
	}
	\label{fig:vort}
\end{figure*} 

Activity can also induce self-organized flow states~\cite{Voituriez2006}. 
In confining channels, an extensile active stress induced instability to bend perturbations leads to self-sustained spontaneous flows~\cite{Voituriez2005,Ramaswamy2007,Edwards2009,Fielding2011,ravnik2013} and spontaneous circulations~\cite{woodhouse2012,jiang2017,theillard2017} when the characteristic activity length scale $\ell_\text{act}$ is comparable to the confinement size $L$, as seen in numerous experimental systems~\cite{wioland2013,lushi2014,beppu2017,Norton2018,Hamby2018,Opathalage2019,Hardouin2020,Thijssen2021}. 
Vortex lattices arise when the characteristic length scale matches the confinement, which can be due to an array of obstacles or cavities~\cite{thampi2016,wioland2016,wioland2016b,nishiguchi2018,reinken2020,james2020}, bound to spherical surfaces~\cite{khoromskaia2017}, or channel confinement~\cite{Shendruk2017,Doostmohammadi2017,zhang2020,samui2021flow}. 
Local circulations arise because $\ell_\text{act}$ represents the characteristic vorticity length scale~\cite{Hemingway2016}, associated with a peak in the enstrophy distribution~\cite{Wensink2012,Bratanov2015,Carenza2020,Alert2020}, and a lattice of vortices represents stress-free solutions that lie at the interface of the stable
and unstable modes in minimal continuum models of active fluids~\cite{slomka2017b}. 
Such 2D flow states represent the emergence of a broken translational symmetry of the vorticity field. 

However, 3D fluids possess the capacity for flows that are not allowed in 2D. 
{\em Helicity}, for instance, is identically zero in 2D and has not previously been observed to spontaneously break symmetry in response to activity.
One reason for this apparent absence is simply because much of the theoretical, computational and experimental work on active nematics is on 2D films. 
However, the recent advent of experimentally realizable 3D active nematics~\cite{duclos2020} has spurred immediate interest in active liquid crystals beyond films~\cite{Shendruk2018,Copar2019,Chandrakar2020,Chandragiri2020,Binysh2020,ruske2021,varghese2020,houston2021}. 
By removing the dimensional limitations of confinement in 2D, additional classes of self-organized flows states may be possible. 
We report long-lived, non-zero helical active dissipative 3D structures that spontaneously break chiral symmetry and have not to our knowledge been previously been observed. 

We numerically confine active nematics in square ducts of size $L$ defined by four impermeable, no-slip walls. 
The square-duct geometry is utilized to produce quasi-1D confinement without curved boundaries~\cite{Ellis2018,Pearce2019,maroudas2021}. 
We vary the extensile activity, reduced temperature, and anchoring strength on the channel walls. 
When the anchoring is weak and reduced temperature is high, we find a {\em double helix} flow structure, in which two streams of anti-parallel flow braid around one another spontaneously breaking chiral symmetry (\fig{fig:flows}a; \movie{mov:DH}). 
When the anchoring strength is strong, we find a {\em grinder train} flow structure, in which a lattice of axially aligned, counter-rotating vortices drift down the channel with non-zero helicity (\fig{fig:flows}b; \movie{mov:GT}). 
Both flow states are helical, span the entire system and are long-lived. 
The self-organization of helicity from a system with nematic symmetry highlights the potential of activity as a pathway for designing emergent, material dynamics. 
Spontaneous chiral symmetry breaking is characteristic of helix formation, which extends beyond active fluids. 

In our model of an active nematic fluid, the non-dimensionalized velocity $\vec{u}(\vec{r}; t)$ and nematic order $\tens{Q}(\vec{r}; t)$ are coupled, and density is constant~\cite{Marenduzzo2007,thampi2014a,Rivas2020}. 
The dimensionless Cauchy equation balances the material derivative of velocity against the divergence of the stress, $\tilde{\rho} D_t \vec{u} = \del\cdot\tens{\Pi}$. 
Since the active contribution to the stress $-\zeta\tens{Q}$ drives spontaneous nematic distortions on length scales $\ell_\text{act}$ and flows of speed $U$, we non-dimensionalize each term by the activity $\zeta$, resulting in the active inertia number $\tilde{\rho} = \rho U^2 / \zeta$ for density $\rho$. 
Two additional dimensionless numbers appear from the stress divergence --- dimensionless viscosity (\eqsi{eq:nondimEta}) and, more relevantly, the  {\em activity number}  $\tilde{K}^{-1/2}=L/\ell_\text{act}$, where $\ell_\text{act}\sim\sqrt{K/\zeta}$ is the characteristic activity length due to the competition of activity $\zeta$ and nematic elasticity $K$.

The evolution equation for the nematic orientation $D_t \tens{Q} - \tens{\mathcal{S}} = \tilde{\Gamma} \tens{\mathcal{H}}$, in which the covariant derivative (including a co-rotational advection term $\tens{\mathcal{S}}$), is balanced by the relaxation towards equilibrium, characterized by an inverse nematic P\'{e}clet number $\tilde{\Gamma}$. 
The molecular field $\tens{\mathcal{H}}$ has contributions due to the bulk (characterized by the lowest order Landau free energy coefficient $A$), distortions (by $K$) and surface anchoring (by anchoring strength $W$). 
Non-dimensionalizing these by the distortion free energy scale produces two dimensionless numbers --- the distance from the passive-thermodynamic isotropic-nematic transition, which is a reduced temperature $\tilde{T}^{1/2} = \sqrt{A/K} L$; and the strength of the degenerate planar anchoring, which can be measured as $\tilde{W}=WL/K$, as in colloidal liquid crystals~\cite{Aryasova2004}. 
Further technical details of the model are provided in \appndx{app:methods}. 

Non-dimensionalized parameter space is denoted by 
\begin{align}
    \tilde{P} &= \left\{ \tilde{K}^{-1/2},\tilde{W}, \tilde{T}^{1/2} \right\} 
        \equiv \left\{ \sqrt{\frac{\zeta}{K}} L, \frac{W}{K}L, \sqrt{\frac{A}{K}}L \right\}, 
\end{align}
each component of which can be interpreted as a ratio of system size to a characteristic length scale (\appndx{app:nonDimScale}). 
By varying anchoring strength $\tilde{W}$ and reduced temperature $\tilde{T}^{1/2}$, the simulations explore regions of parameters space that have previously not received adequate attention. 


For sufficiently small extensile activity ($\tilde{K}^{-1/2} \lesssim 18$), we find spontaneous streaming in the axial direction. 
These are unidirectional and oscillating flows, which have been well documented in 2D channels~\cite{giomi2008,Fielding2011,Shendruk2017}. 
Here, we do not delineate between these but refer to both as {\em axial streaming}. 
At sufficiently large activity, mesoscale or {\em active turbulence} occurs. 
In active turbulence, the components of vorticity $\vec{\omega}=\del\times\vec{u}$ in the axial and two transverse directions (denoted by subscript $\parallel$ and $\perp_{1,2}$ respectively) are isotropic---the probability distributions of the axial $\omega_\parallel$ and transverse $\omega_{\perp_{1,2}}$ components of the vorticity are equivalent (\fig{fig:vort}d). 
The distributions are symmetric about zero and normal. 

The active dissipative structures exist in the intermediate regime between axial streaming and active turbulence, where the active length scale $\ell_\text{act}$ is comparable to the confinement length $L$. 
The most common of these dissipative structures is a non-helical lattice of counter-rotating vortices oriented transverse to the channel, breaking translational symmetry (\fig{fig:flows}c). 
This is the 3D equivalent of the 2D Ceilidh dynamic state~\cite{Shendruk2017,Doostmohammadi2017}. 
The 3D Ceilidh lattice exhibits dancing disclinations; however, these are now curved disclination lines that span the channel (\movie{mov:disc}). 
As a vortex lattice, the vorticity distribution is strongly bimodal (\fig{fig:vort}c). 
The $\omega_{\perp_{2}}$ distribution is symmetric about zero and bimodal with prominent peaks, representing the lattice of counter-rotating vortices with spontaneous symmetry breaking between the two transverse directions. 
There is negligible vorticity in both the other directions.

\begin{figure}[tb]
	\centering
	\includegraphics[width=0.5\textwidth]{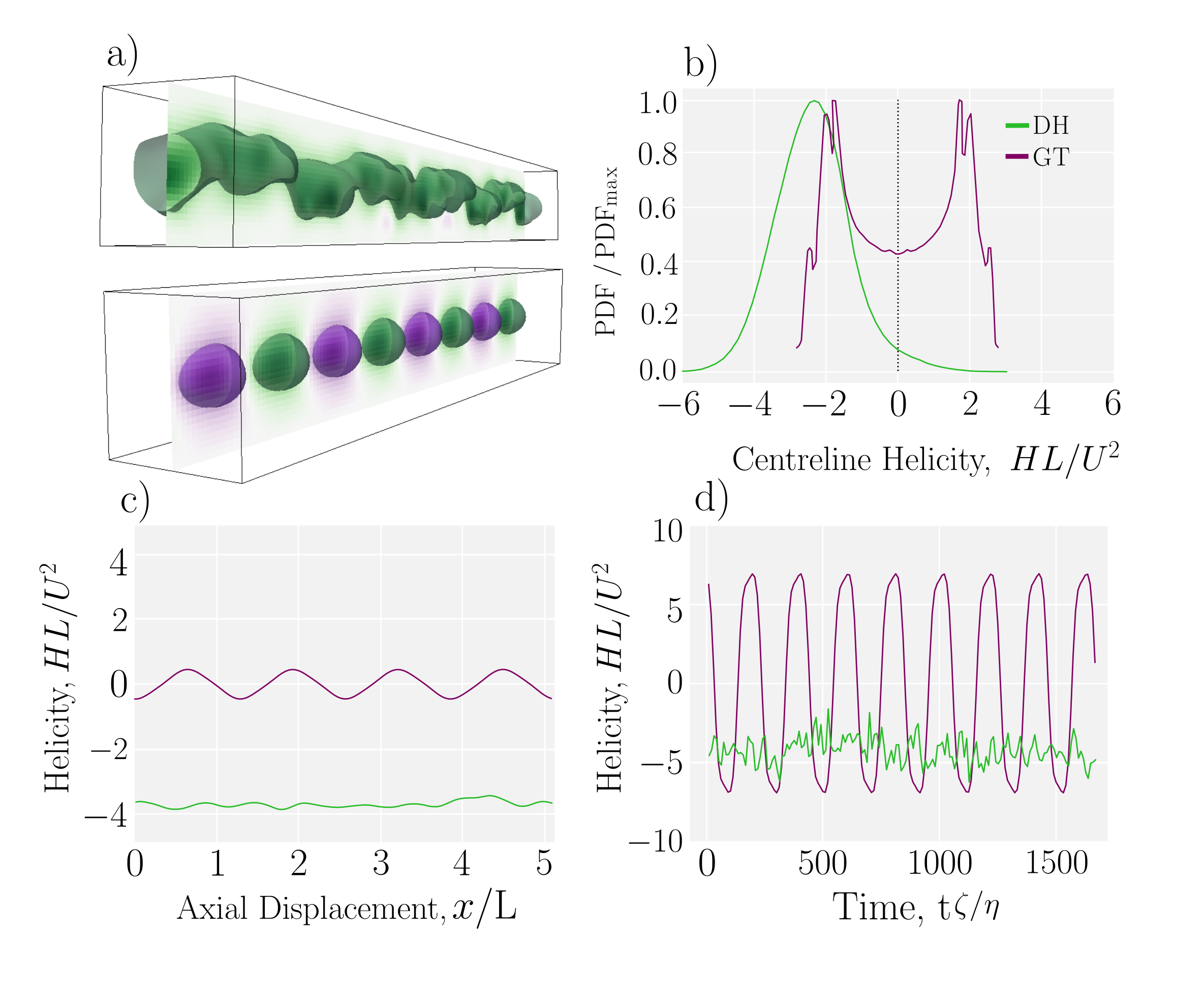}
	\caption{\small
	\textbf{Local helicity} $H=\vec{u}\cdot\vec{\omega}$ for the double helix ($\tilde{P}=\{25,0.0015,6.5\}$) and grinder train ($\tilde{P}=\{25,250,0\}$ flow states). 
    (a) Snapshots showing the helicity field as a 3D isosurface through the center plane of the duct.
    \textbf{Top:} Double helix with negative chirality (\movie{mov:h-DH}). 
    \textbf{Bottom:} Grinder train with alternating chirality (\movie{mov:h-GT}). 
    \textbf{(b)} Distributions of helicity down the channel centreline averaged over ten random initializations.
    \textbf{(c)} Spatial variation of instantaneous cross-sectional averaged helicity.
    \textbf{(d)} Temporal variation of cross-sectional averaged helicity at a single point. 
    }
	\label{fig:helicity}
\end{figure}

While the grinder train is similar to the Ceilidh lattice, the crucial distinction is that the counter-rotating vortices are oriented axially, rather than transversely (\fig{fig:flows}b; \movie{mov:GT}). 
Like axial streaming states~\cite{Voituriez2006,Chandragiri2020}, the grinder train has a net flow along the channel. 
This flow structure is a train of axially aligned counter-rotating vortices drifting down the axis of the channel that exists when anchoring is strong (\movie{mov:GT}). 
Thus, the distributions of vorticity are similar to the Ceilidh vortex, except the bimodal distribution is narrower and in the axial direction $\omega_\parallel$, rather than a transverse direction (\fig{fig:vort}b). 
Though both the Ceilidh lattice and the grinder train manifest counter-rotating vortices and both break translational symmetry, the Ceilidh lattice does not possess local helicity $H=\vec{u}\cdot\vec{\omega}$ in contrast to the grinder train (\fig{fig:helicity}a). 
The instantaneous centreline helicity forms a well-defined wave (\fig{fig:helicity}c) and the drift exhibits temporal oscillation (\fig{fig:helicity}d; \movie{mov:h-GT}), reflecting  the steady motion of the helicity train. 
The steady translation, but oscillating sign, of the helicity indicates the grinder is a lattice of counter-rotating vortices. 

Qualitatively, the grinder train possesses characteristics of both axial streaming and vorticity translational symmetry breaking. 
Indeed, as the axial streaming begins to oscillate on scales comparable to the confinement, counter-rotating vortices can align axially, resulting in non-zero local helicity---though the average is zero (\movie{mov:h-DH}), the distribution is bimodal (\fig{fig:helicity}b). 
It also generates smaller vorticity (\fig{fig:vort}b-c) due to the fact that it possesses less nematic distortions and is entirely free of disclinations (\movie{mov:disc}). 

\begin{figure*}[tb]
	\centering
	\includegraphics[width=0.99\textwidth]{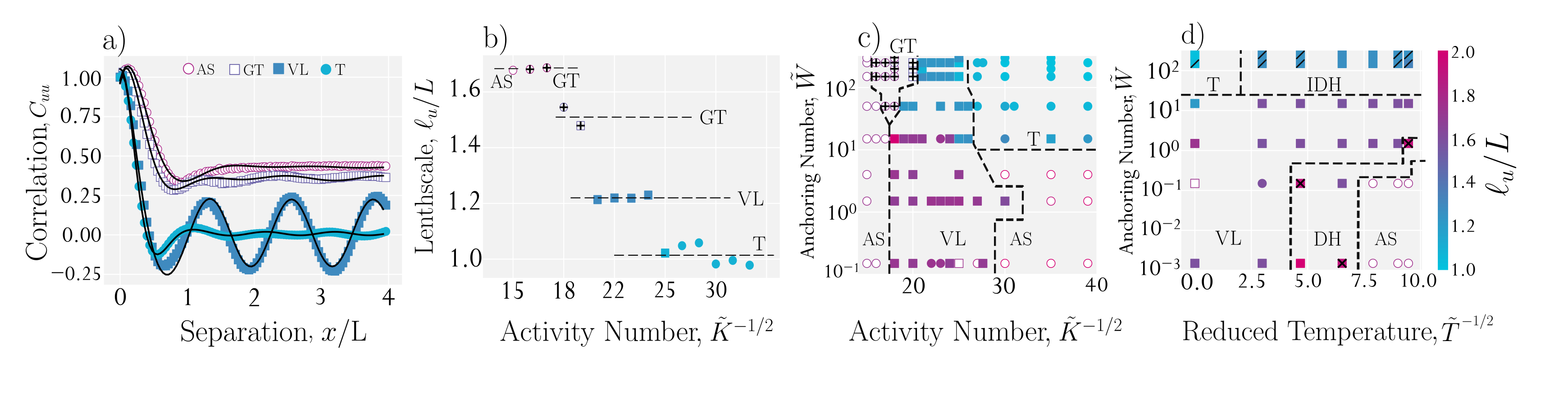}
	\caption{\small
	\textbf{Active dissipative structure phase diagrams.} 
 	\textbf{(a)} Spatial velocity-velocity correlations along the channel centre line for $\tilde{P} = \{ \tilde{K}^{-1/2},250,0 \}$ for various $\tilde{K}^{-1/2}$. 
 	Example curves for axial streaming (AS; $\tilde{K}^{-1/2}=15$), grinder train (GT; $\tilde{K}^{-1/2}=20$), vortex lattice (VL; $\tilde{K}^{-1/2}=15$) and mesoscale turbulence (T; $\tilde{K}^{-1/2}=15$). 
 	Open symbols denote non-zero long-range correlations, while square symbols denote long-range oscillations with non-zero amplitude. 
 	\textbf{(b)} Characteristic length scale $\ell_u$ determined from the autocorrelation functions (\appndx{app:lengths}) for $\tilde{P} = \{ \tilde{K}^{-1/2},250,0 \}$. 
 	 $\boxplus$ marks large values of peak resolution (\eqsi{eq:res}).
 	\textbf{(c)} The length scales $\ell_u$ as a function of dimensionless extensile activity $\tilde{K}^{-1/2}$ and planar anchoring $\tilde{W}$ for reduced temperature $\tilde{T}^{1/2}=0$. 
 	\textbf{(d)} Length scales as a function of $\tilde{W}$ and $\tilde{T}^{1/2}$ for $\tilde{K}^{-1/2}=25$. 
 	Markers denote for average system helicity (\appndx{app:helicityAv}): 
 	$\boxtimes$ denotes large average helicity $H_{\abs{\langle\rangle}}$ and $\squarediv$   indicates 
    large values of $H_{\langle\abs{}\rangle}$, which respectively identify regions for the double helix (DH) and intermittent double helix (IDH). 
	}
	\label{fig:corrLength}
\end{figure*}

On the other hand, the double helix does possess large nematic distortions: 
a clear cork-screw disclination line spirals through the centre of the channel (\movie{mov:disc}). 
This is because the grinder train exists at strong anchoring, while the double helix exists at weak anchoring and high reduced temperature, which leads to more disorderly helicity fields (\fig{fig:helicity}d; \movie{mov:h-DH}). 
The helicity is largest far from the no-slip boundaries. 
We find double helices most frequently in regions of high reduced temperature $\tilde{T}^{1/2}$, suggesting that the double helix emerges when the thermodynamic persistence length is comparable to the confinement size. 
Unlike the Ceilidh or grinder lattices, the double helix is not composed of isolated vortices and so the centreline helicity does not exhibit clear spatio-temporal structure (\fig{fig:helicity}c-d). 
Additionally, the vorticity of the double helix state is smallest in the axial direction and not strongly bimodal in the two transverse directions (\fig{fig:vort}a). 
Unlike the grinder train, the helicity distribution is primarily unimodal (\fig{fig:helicity}b). 
Since it is composed of a pair of helices with opposite handedness but also opposite streaming directions, the double helix spontaneously breaks chiral symmetry. 

To map the existence of helical flow in parameter space, we characterize the flow states by their associated length scales and helicity. 
The non-helical Ceilidh vortex lattice is long-lived and system-spanning, with spatial persistence clear from the velocity-velocity autocorrelation function $C_{uu}$ (\appndx{app:lengths}) down the length of the channel (\fig{fig:corrLength}a; blue squares), exhibiting long-range oscillating correlations. 
Dissimilarly, the correlation function of the turbulent state rapidly decorrelates (\fig{fig:corrLength}a; blue circles). 
While the correlation functions of these two intercept zero (\fig{fig:corrLength}; filled symbols), the axial streaming state initially decays to a long-range constant value (\fig{fig:corrLength}a; open pink circles) and the grinder train possesses long-range oscillating correlations about a non-zero value (\fig{fig:corrLength}a; open purple squares). 
The correlation function of the grinder train has a longer period but smaller amplitude than the Ceilidh lattice and does not decorrelate (\fig{fig:corrLength}a) since the grinder train state appears between axial streaming and the vortex lattice (\fig{fig:corrLength}b). 

The active dissipative structures are identified by their characteristic length scales $\ell_{u}$, measured from the velocity correlations (\appndx{app:lengths}). 
The length scales reveal distinct transitions between each flow state (\fig{fig:corrLength}b). 
The Ceilidh lattice occupies the largest region of the activity-anchoring phase plane (\fig{fig:corrLength}c). 
The phase diagram demonstrates that axial streaming exists for all anchoring numbers at sufficiently small activity. 
The activity at which the flow transitions from axial streaming is only weakly dependent on anchoring strength across many orders of magnitude (\fig{fig:corrLength}c). 
For moderate to weak anchoring, the Ceilidh lattice is the only dissipative structure between axial streaming and active turbulence. 

The grinder train arises because the anchoring suppresses the formation of the dancing disclination lines that are present in the vortex lattice. 
This reduction in nematic distortion lowers the vorticity magnitude and makes the flow deterministic, such that the transverse directions coincide (\fig{fig:vort}b). 
However, it does not alter the match between the characteristic vortex size $\ell_\text{act}$ and the confinement $L$. 
Thus, the vortices preferentially lie axially along the channel.

When the reduced temperature $\tilde{T}^{1/2}$ is increased above the isotropic-nematic transition, the fluid can still be dynamically oriented by the active flows~\cite{santhosh2020,Giordano2021}. 
Indeed, vortex lattices exist for a wide range of $\tilde{T}^{1/2}$ and $\tilde{W}$ at $\tilde{K}^{-1/2}=25$ (\fig{fig:corrLength}d). 
However, in the low anchoring limit and at moderate reduced temperatures, the double helix structure exists as a second region for which there is non-zero helicity. 
While the grinder train and double helix both exist for intermediate activity, the grinder train requires strong anchoring at $\tilde{T}^{1/2}=0$ (\fig{fig:corrLength}c) and the double helix occurs for weak anchoring at high reduced temperatures (\fig{fig:corrLength}d).

In addition to the orderly double helix and grinder train, we identify a less structured, noisy regime at high reduced temperatures and strong anchoring (\fig{fig:corrLength}d; slashes). 
In this regime, we find intermittent double helices interspersed with more chaotic behaviour (\movie{mov:IDH}). 
Naturally, the chaos in these systems results in inconsistent helicity but higher fluid speeds, resulting in lower average helicity. 

To understand the origin of active helical structures, we derive the transport equation for helicity
\begin{align}
 \tilde{\rho} D_t H +\del \cdot \vec{J} &= \Sigma   
\end{align}
with helicity flux $\vec{J}$ and source terms $\Sigma$ (see \appndx{app:helicity}). 
By calculating the dimensionless numbers associated with each source (\eqssi{eq:sources}-\eqref{eq:actSources}), we find that the nematic and viscous sources are  negligible compared to the activity-induced helicity source $\Sigma^\text{a} = -\vec{\omega} \cdot [\del\cdot\tens{Q}] - \vec{u} \cdot (\del\times [\del\cdot\tens{Q}] )$.
The first term emerges from the projection of active force on to vorticity field, while the second term represents projection of the curl of the active force onto the velocity field. 
In both helical flow states, the two active terms are of the same magnitude (\figsi{fig:helicitySources}). 
Immediately following initialization, the transient grinder train structure begins as oscillating discoidal regions of $\Sigma^\text{a}$ (\movie{mov:GT-actSource}). 
Once the drifting grinder train forms, $\Sigma^\text{a}$ takes the form of a travelling wave that produces the drifting train of alternating helicity (\fig{fig:helicity}). 
Similarly for the double helix, the active source forms local regions of production and elimination (\movie{mov:DH-actSource}), which are associated with the cork-screw disclination. 

Recent work on 3D active nematics~\cite{Shendruk2018,Copar2019,Chandrakar2020,Chandragiri2020,Binysh2020,ruske2021,varghese2020,houston2021} has shown them to be complicated by their tangle of 3D disclination lines.  
However, they also possess the potential for an exciting range of possible dissipative structures. 
We have shown a simple case with the emergence of two previously unobserved spontaneous, long-lived, non-zero helicity structures in simple confinement for extensile activity, planar anchoring and non-negative reduced temperature. 
In driven systems, double helical flow states can exhibit complex non-linear dynamics~\cite{Delbende2021} and control the dispersivity of particles eluting in streams~\cite{song2009}. 
In fact, surface-activity driven helical flows have been identified and analyzed in the context of cytoplasmic streaming~\cite{Goldstein2008,Meent2008,verchot2010,woodhouse2013}. 
In these cases of long plant cells, helical flow enables significant transport, mixing and enhanced rates of nutrient exchange with the surrounding membrane~\cite{Goldstein2008}. 
Indeed, recent work has considered defect dynamics~\cite{Pearce2020} and helical flows~\cite{Napoli2020} on cylindrical surfaces. 
Spontaneous chiral symmetry breaking is a characteristic aspect of helix formation in such material dynamics, which extends beyond the sphere of active fluids. 
For example, chirality has been shown to be relevant in morpohological processes, such as the gastrulation of \textit{Drosophila} embryos, in which gut formation acquires spontaneous left-right asymmetry that leads to twisting~\cite{Hozumi2006}. 
Likewise, the mitotic spindle, which has been modelled as an active nematic~\cite{Oriola2020}, exhibits chirality~\cite{Novak2018}. 
Furthermore, spontaneous chiral symmetry breaking has been observed in the director field of passive achiral nematics subject to strong confinements~\cite{tortora2011, Koning2014, nayani2015, jeong2015}. 

The experimental realization of 3D active nematics~\cite{duclos2020} promises many opportunities to explore active dissipative structures that were not possible in 2D and helicity may play a key in many such structures, as suggested by these results. 
Helicity is involved in energy cascades in active turbulence~\cite{slomka2017a} and generally plays an important role in characterizing the topological nature of 3D flows~\cite{Moffatt2021}. 
This letter reports spontaneous chiral symmetry breaking in active nematics leading to a steady state helical flow. 
Indeed, in stark contrast to the defect dancing observed in the Ceilidh vortex lattice state, the grinder train does not possess any topological singularities, while the double helix possesses a single disclination line that winds through the two helical flows. 

\section*{Acknowledgements}
We thank Kevin Stratford and Kristian Thijssen for invaluable help. 
This research has received funding (TNS) from the European Research Council under the European Union’s Horizon 2020 research and innovation programme (Grant agreement No. 851196). 
SPT acknowledges the support by the Department of Science and Technology, India, via the research grant CRG/2018/000644. 
TA-N has been supported in part by the Academy of Finland PolyDyna (no. 307806) and QFT Center of Excellence Program grant (no. 312298).


\clearpage
\appendix

\section{Numerical Methods}
\label{app:methods}
Simulations numerically solve the incompressible active nematohydrodynamics equations of motion~\cite{Marenduzzo2007,thampi2014a} for the local non-dimensionalized velocity $\vec{u}(\vec{r}; t)$ and nematic order $\tens{Q}(\vec{r}; t)$, for constant density, $\del\cdot\vec{u}=0$. 
The non-dimensionalized Cauchy equation balances the material derivative $D_t=\partial_t+\vec{u}\cdot\del$ of velocity against the divergence of the stress $\tens{\Pi}$ 
\begin{align}
    \tilde{\rho} D_t \vec{u} &= \del\cdot\tens{\Pi}
    \label{eq:mom}. 
\end{align}
Since activity drives the fluid flow, the stress is normalized by the magnitude of the activity $\zeta$, resulting in a dimensionless {\em active inertia number}
\begin{align}
    \label{eq:nondimRho}
    \tilde{\rho} &= \frac{\rho U^2}{\zeta},
\end{align}
which compares inertial to active forces in an active fluid of density $\rho$ and spontaneous characteristic speed $U$. 
The non-dimensionalized stresses in \eq{eq:mom} have four contributions~\cite{denniston2004}: 
\begin{description}
    \item[\textit{(i)}] active stress $\tens{\Pi}^\text{a}=-\tens{Q}$; 
    \item[\textit{(ii)}] viscous stress $\tens{\Pi}^\text{v}=2\tilde{\eta} \tens{E}$, where $2\tens{E}=\del \vec{u}+ \left(\del\vec{u}\right)^\intercal$ is the strain rate and 
    \begin{align}
        \label{eq:nondimEta}
        \tilde{\eta} &= \frac{\eta U}{\zeta L}
    \end{align}
    is the dimensionless viscosity number; 
    \item[\textit{(iii)}] nematic contribution $\tens{\Pi}^\text{n}=\tilde{K}\tens{\pi}^\text{n}$, where $\tens{\pi}^\text{n} =2\lambda(\tens{Q}:\tens{\mathcal{H}})\tens{\mathcal{Q}} - \lambda \left\{\tens{\mathcal{H}},\tens{\mathcal{Q}}\right\}_{+} + \left\{\tens{Q},\tens{\mathcal{H}}\right\}_{-} - \del \left( \tens{Q}: \tfrac{\delta F}{\delta\del\tens{Q}} \right)$ in which $\tens{\mathcal{Q}}=\tens{Q}+\tens{\delta}/3$, $\tens{\mathcal{H}} = -(\frac{\delta F}{\delta\tens{Q}} - \frac{1}{3}\tens{I} \; \text{Tr}\frac{\delta F}{\delta\tens{Q}})$ is the molecular field, $\lambda$ is an alignment parameter, and $\left\{\tens{A},\tens{B}\right\}_\pm = \tens{A}\cdot\tens{B} \pm \tens{B}\cdot\tens{A}$ is a (anti)commutator; and
    \item[\textit{(iv)}] hydrodstatic pressure $\tens{\Pi}^\text{p}=-P \tens{\delta}$, where pressure is non-dimensionalized by activity. 
\end{description}
The dimensionless number
\begin{align}
    \label{eq:nondimFrank}
    \tilde{K} &= \frac{K}{\zeta L^2}
\end{align}
assumes a single elastic Frank coefficient $K$ and is the squared ratio of the characteristic activity length in bulk $\ell_\text{act}\sim\sqrt{K/\zeta}$ and the scale of confinement $L$. 
A convenient dimensionless activity number is then 
\begin{align}
    \tilde{K}^{-1/2} &= \frac{L}{\ell_\text{act}}, 
\end{align}
which is employed in the main text as the dimensionless measure of activity. 

The transport equation for the nematic order tensor balances the relaxation towards equilibrium against the generalized material derivative 
\begin{align}
  D_t \tens{Q} -\tens{\mathcal{S}} &= \tilde{\Gamma} \tens{\mathcal{H}} \label{eq:nem}, 
\end{align}
including co-rotational advection, 
\begin{align}
    \label{eq:corot}
    \tens{\mathcal{S}} &= \left\{ \lambda \tens{E}+\tens{\mathcal{W}},\tens{\mathcal{Q}} \right\}_+ - 2\lambda\left(\tens{Q}:\del\vec{u}\right)\tens{\mathcal{Q}} ,
\end{align}
where $2\tens{\mathcal{W}} = \left(\del\vec{u}\right)^\intercal - \del \vec{u}$ is the vorticity tensor. 
An {\em inverse nematic P\'{e}clet number} 
\begin{align}
    \label{eq:nondimPeclet}
    \tilde{\Gamma} &= \frac{\Gamma K}{LU}
\end{align}
relates elastic rotational relaxation rate $\Gamma K/L^2$ to advection rate $L/U$. 

\begin{figure}[tb]
	\centering
	\includegraphics[width=0.5\textwidth]{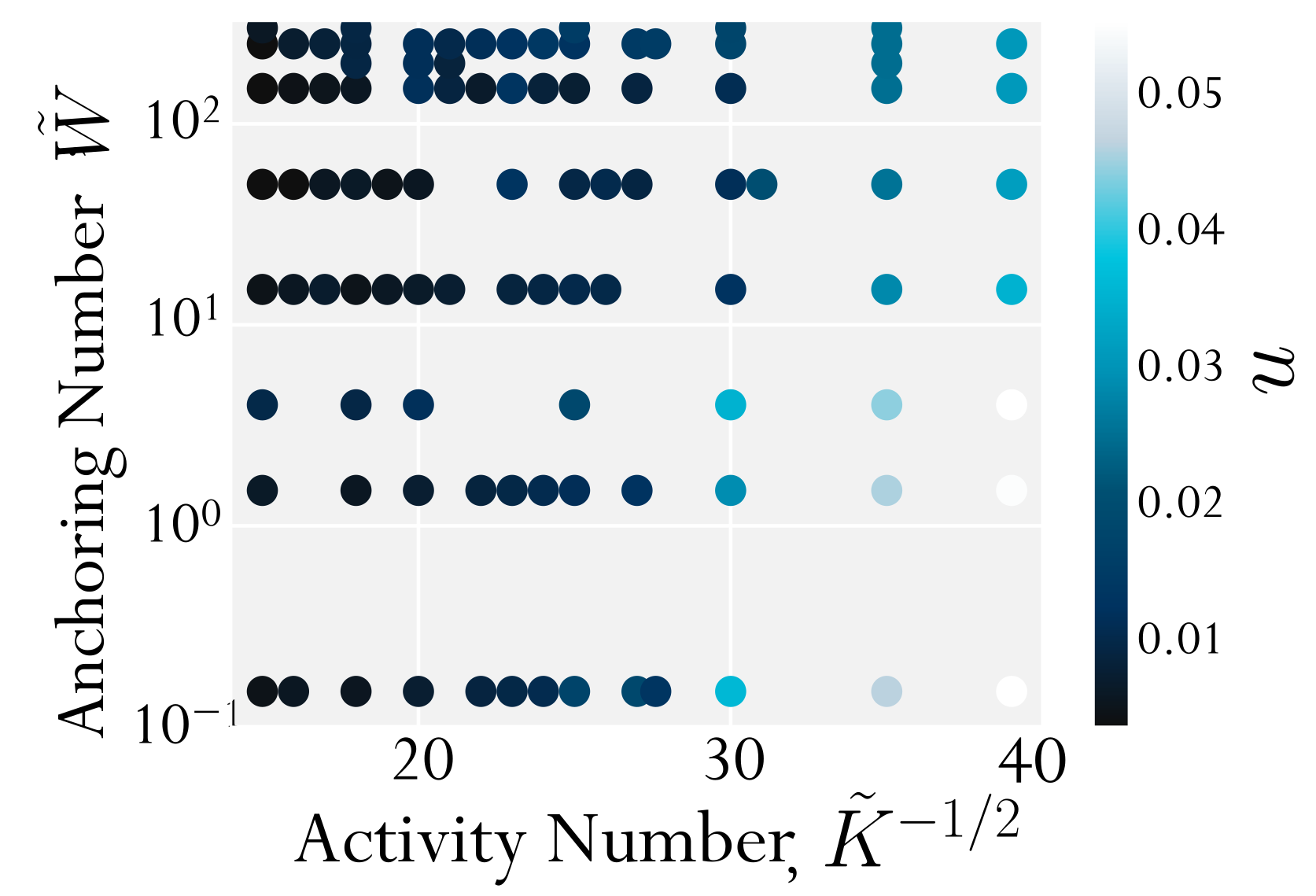}
	\caption{
	Average root mean squared velocities $U$ in LB simulation units for reduced temperature $\tilde{T}^{1/2}=0$.
	}
	\label{fig:U}
\end{figure}

The non-dimensionalized molecular field $\tens{\mathcal{H}} = -\tfrac{\delta F}{\delta \tens{Q}} + \tfrac{\tens{I}}{3} \text{Tr}\left[\tfrac{\delta F}{\delta \tens{Q}}\right]$ appears in both \eq{eq:mom} and \eqref{eq:nem} and belongs to the free energy $F$, which is comprised of Landau-de~Gennes and deformation contributions 
$F= \tilde{A}_0 \int d^3r \ \tfrac{\left(1-\gamma/3\right)}{2}\tens{Q}:\tens{Q} - \tfrac{\gamma}{3} \tens{Q}:(\tens{Q}\cdot\tens{Q}) + \tfrac{\gamma}{4} \left(\tens{Q}:\tens{Q}\right)^2 + \left(\del\tens{Q}\right)^2$. 
The bulk free energy density has been normalized by the elasticity $\tilde{A}_0=A_0 L^2/K$ and includes the dimensionless parameter $\gamma$, which determines the magnitude of the nematic order with $\gamma=2.7$ the coexistence point between isotropic and nematic phases~\cite{Marenduzzo2007}. 
Within the nematic phase, the characteristic defect core size can be approximated by the thermodynamic nematic persistence length $\ell_Q \sim \sqrt{K/A}$ for $A=A_0\left(1-\gamma/3\right)$~\cite{Hemingway2016}, such that we can define a reduced temperature in terms of a squared ratio of length scales
\begin{align}
    \tilde{T} 
    &= \left(\frac{L}{\ell_Q}\right)^2 
    = \tilde{A}_0 \left(\frac{3-\gamma}{3}\right)
    .
    \label{eq:nondimTemp}
\end{align}
This is convenient, as it allows us to directly compare the effects of thermodynamic nematic ordering to confinement. 
We utilize $\tilde{T}^{1/2}$ as a measure of reduced temperature in the main text. 

In addition to the bulk, the walls impose degenerate planar anchoring~\cite{Fournier2005} via a surface free energy density 
\begin{align} \label{eq:planar_anchoring}
    f_\text{wall} &= W_1\left(\tens{Q}-\tens{Q}^\perp\right):\left(\tens{Q}-\tens{Q}^\perp\right) \nn\\
            &\qquad + W_2\left(\tens{Q}:\tens{Q}-S_\text{eq}^2\right)^2, 
\end{align}
where $6S_\text{eq} = 1+3\sqrt{1-8/(3\gamma)}$ is the equilibrium value of the scalar order parameter. 
The projection of the orientation tensor onto the surface is $\tens{Q}^\perp=\tens{P}\cdot\tilde{\tens{Q}}\cdot\tens{P}$, where $\tilde{\tens{Q}}=\tens{Q}+S_\text{eq}\tens{\delta}/3$, the projection operator is $\tens{P}=\tens{\delta} - \vec{n}_\perp \vec{n}_\perp$ and the surface normal is $\vec{n}_\perp$. 
We set both the surface anchoring constants for the orientation and ordering to the same value  $W=W_1=W_2$ and non-dimensionalize the anchoring
\begin{align}
    \label{eq:nondimAnchoring}
    \tilde{W} &= \frac{WL}{K}. 
\end{align}
As with activity number $\tilde{K}^{-1/2}$ and the reduced temperature $\tilde{T}^{1/2}$, $\tilde{W}$ can be interpreted as the ratio of confinement size $L$ to the de Gennes-Kleman extrapolation length $\ell_\text{dGK} \sim K/W$~\cite{suh2019}. 
In the main text, $\tilde{W}$ serves as the measure of anchoring strength. 

\begin{figure}[tb]
	\centering
	\includegraphics[width=0.5\textwidth]{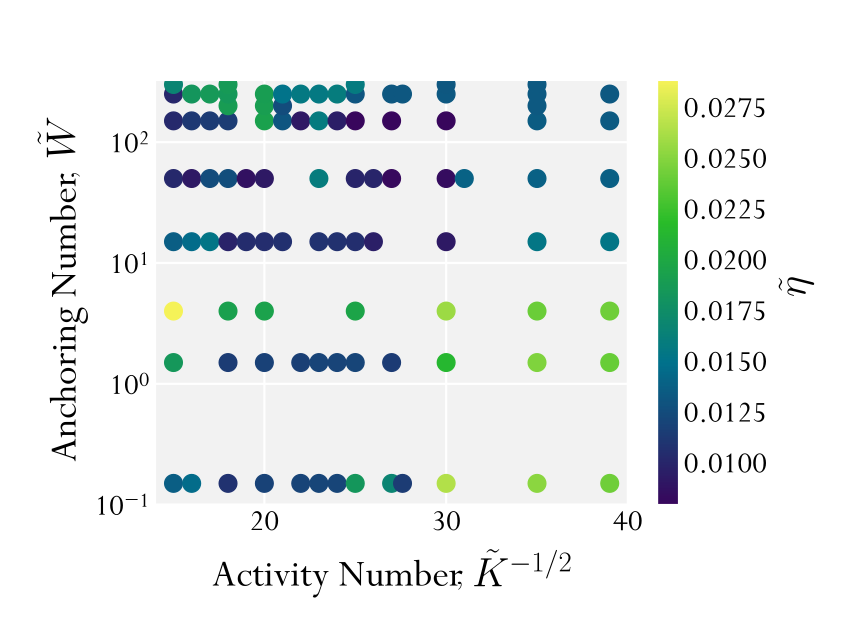}
	\caption{
	Non-dimensional viscosity number,  $\tilde{\eta} = (\eta U)(\zeta L) \sim 10^{-2}$, for reduced temperature $\tilde{T}^{1/2}=0$. 
	}
	\label{fig:eta}
\end{figure}

The equations of motion are solved using a hybrid scheme~\cite{Marenduzzo2007}, in which the Cauchy equation (\eq{eq:mom}) is solved using the lattice Boltzmann algorithm, while \eq{eq:nem} is solved using a finite difference predictor-corrector algorithm. 
Simulations are performed in 3D square channels of size $L=25$ defined by impermeable, no-slip walls and periodic boundary conditions separated by a channel length of $\mathcal{L}\in\{128,200\}$. 
Simulations are run for a minimum of $2\times10^{5}$ time steps to a maximum of $5\times10^{5}$. 
The density $\rho=1$, rotational diffusivity $\Gamma=0.3375$ and alignment parameter $\lambda = 1$ are all held constant. 
On the other hand, Landau-de~Gennes energy density scale $A_0\in\left\{0.1,1\right\}$, distance from the isotropic-nematic transition $\gamma\in[2.668,3]$, dynamic viscosity $\eta=4/3$, Frank elasticity $K\in\left\{0.05,0.05555\right\}$, anchoring energy $W\in\left[3\times10^{-6},0.5\right]$ and activity $\zeta\in\left[0.018,0.122\right]$ are varied throughout.

\section{Magnitude of non-dimensional numbers}
\label{app:nonDimScale}

Not all the terms in the active nematohydrodynamic equations of motion are significant. 
The parameters produce flows with average root mean squared velocities $U\sim10^{-2}$ (\fig{fig:U}). 
The lowest values ($U\approx3.4\times10^{-3}$) occur in the vortex lattice phase (intermediate $\tilde{K}^{-1/2}$ for low anchoring $\tilde{W}$), as well as axial streaming for intermediate anchoring $\tilde{W}$. 
The largest ($U\approx5.5\times10^{-2}$) occur for high activities and low anchoring --- in which the fluid streams at high velocities with non-zero long-range correlations. 
Throughout the main text, we take $U$ to be the characteristic velocity scale and employ it to normalize the quantities of interest. 

\begin{figure}[tb]
	\centering
	\includegraphics[width=0.5\textwidth]{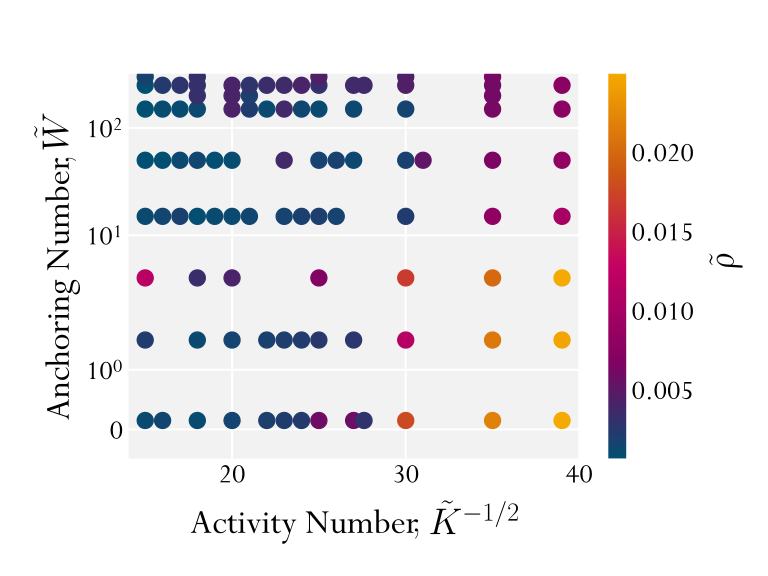}
	\caption{
	Non-dimensional inertia number, $\tilde{\rho} = \rho U^2/\zeta \sim 10^{-2}$, for reduced temperature $\tilde{T}^{1/2}=0$. 
	}
	\label{fig:rho}
\end{figure}

The small values of $U$ cause many of the non-dimensionless numbers from \appndx{app:methods} to be small. 
In particular, the dimensionless viscosity number $\tilde{\eta}=(\eta U)/(\zeta L)$ (\fig{fig:eta}) can be interpreted as a ratio of the measured spontaneous flow speed in the channel in units of the characteristic active speed $\zeta L/\eta$. 
Because of this, the map of $\tilde{\eta}$ is identical to the map for $U$ (\fig{fig:U} and \fig{fig:eta}). 
The fact that $\tilde{\eta}\sim10^{-2}$ highlights the difference between confined active nematics and unconfined, bulk active turbulence: 
In bulk, one expects that the spontaneous flow directly results from the balance viscous forces to active force and so $\tilde{\eta}\sim1$. 
However, the no-slip boundaries modify this expectation and cause $\tilde{\eta}$ to be substantially smaller. 

The dimensionless inertia number $\tilde{\rho} = \rho U^2/\zeta$ (\fig{fig:rho}) is qualitatively similar to the maps of $\tilde{\eta}$ and $U$. 
Indeed, the order of magnitude of $\tilde{\rho} \sim 10^{-2}$ is the same as the dimensionless viscosity $\tilde{\eta}$.  
Because $\tilde{\rho} \sim U^2$, the dimensionless inertia number is much larger in the region with the fastest flows (high activities and low anchoring) than elsewhere. 
The small scale of $\tilde{\rho}$ indicates that hydrodynamic flows that result from \eq{eq:mom} is primarily a balance of stresses $\del\cdot\tens{\Pi}\approx0$. 

\begin{figure}[tb]
	\centering
	\includegraphics[width=0.5\textwidth]{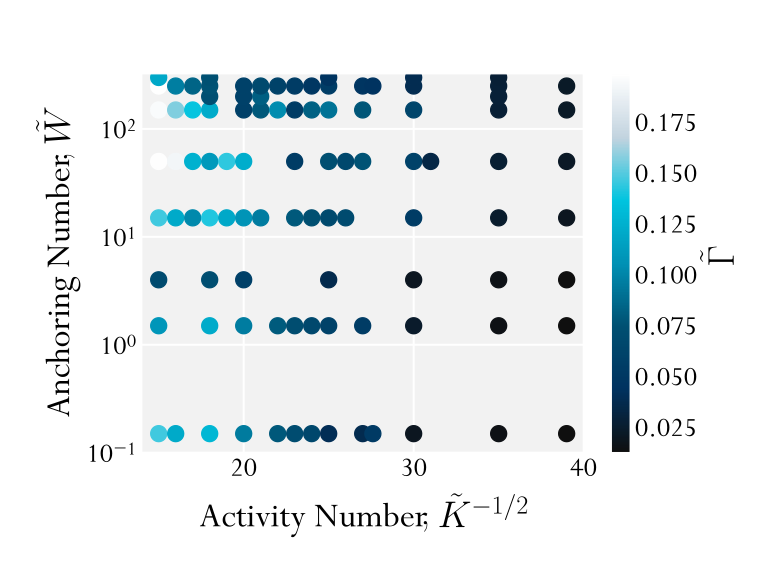}
	\caption{
	Non-dimensional inverse P\'{e}clet number,  $\tilde{\Gamma} = (\eta U) / (\zeta L) \sim 10^{-2}$, for reduced temperature $\tilde{T}^{1/2}=0$.
	}
	\label{fig:gamma}
\end{figure}

The inverse Peclet number $
\tilde{\Gamma} = (\Gamma K)/(LU)$ is also relatively small, with values $
\tilde{\Gamma} \sim 10^{-1}$ (\fig{fig:gamma}), despite $\tilde{\Gamma}\sim U^{-1}$. 
This is because $\gamma$ and $K$ are both small and $L$ is large. 
Because $\tilde{\Gamma}<1$, the generalized material derivative in \eq{eq:nem} is more significant than the thermodynamic relaxation towards equilibrium, which indicates $D_t \tens{Q} \approx \tens{\mathcal{S}}$. 
Thus, the nematic orientational order field are primarily governed by the active flows. 

Thus, the other three dimensionless numbers from \appndx{app:methods} act as our dimensionless parameter space $\tilde{P} = \{ \tilde{K}^{-1/2},\tilde{W},\tilde{T}^{1/2} \}$. 
Each of these numbers can be interpreted as a ratio of length scales. 
We consider dimensionless activity numbers $\tilde{K}^{-1/2}=L/\ell_\text{act}$ in the range $10$ to $40$, which is the regime in which flows transition from axial streaming to mesoscale turbulence. 
The anchoring number $\tilde{W}=L/(K/W)$ varies over five orders of magnitude, from $10^{-3}$ to $250$. 
The reduced temperature $\tilde{T}^{1/2}=L/\ell_Q$ is varied from $0$ (representing the thermodynamic isotropic-nematic transition point) to $10$ (well within the isotropic phase). 

\begin{figure*}[tb]
	\centering
	\includegraphics[width=0.95\textwidth]{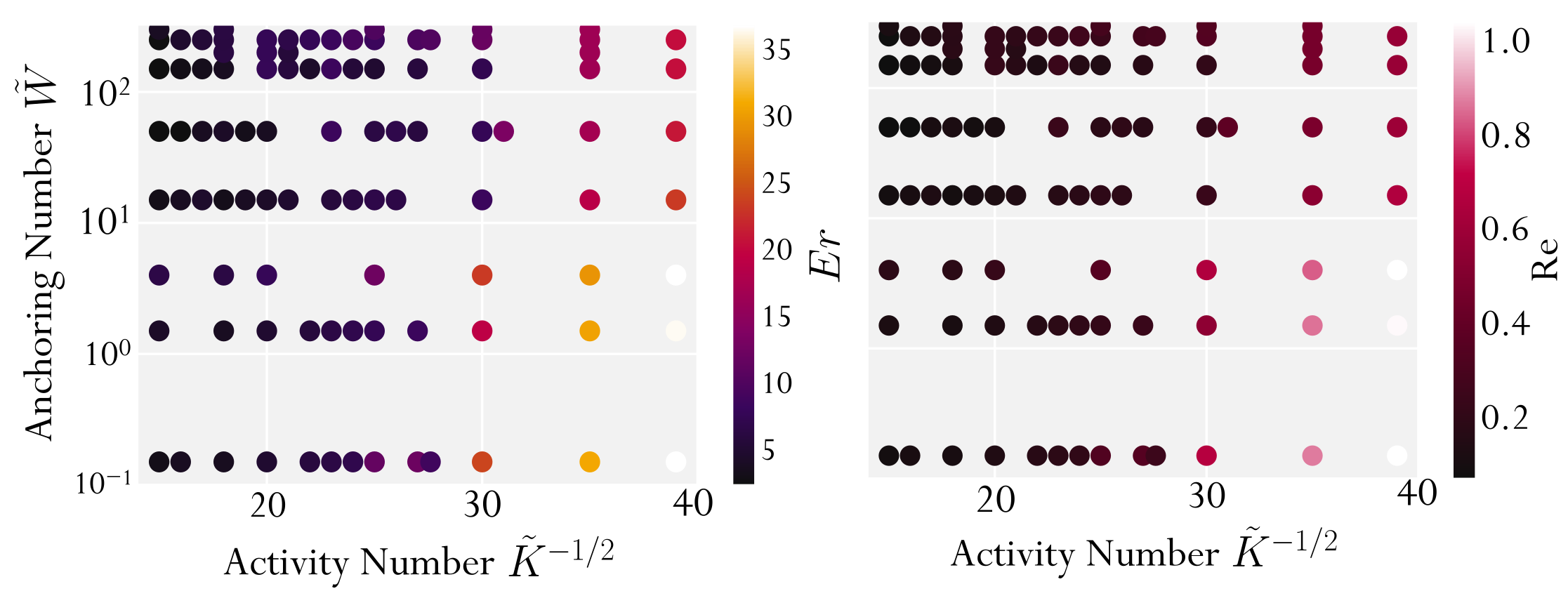}
	\caption{ Ericksen numbers $\text{Er}=\tilde{\eta}/\tilde{K}$ (left) and Reynolds numbers $\text{Re}=\tilde{\rho}/\tilde{\eta}$ (right) for reduced temperature $\tilde{T}^{1/2}=0$. 
	}
	\label{fig:re_er}
\end{figure*}

While the Reynolds and Ericksen numbers do not explicitly appear, they are both intermediate in magnitude (\fig{fig:re_er}). 
The Ericksen number $\text{Er}=\tilde{\eta}/\tilde{K}$ rises as high as $35$.
However, this is restricted to the high activity/low anchoring region of active streaming. 
In the regions of interest (vorticity lattices, grinder train and double helix), $\text{Er}\sim3$. 
This suggests that active flows are moderately more significant than elastic response. 
Likewise, the Reynolds number $\text{Re}=\tilde{\rho}/\tilde{\eta}$ can approach unity in the high activity region but is $\text{Re}\sim0.1$ in the regions of interest. 
This indicates that viscous forces are somewhat more prominent than inertial ones. 
However, as described above, $\tilde{\rho}$ is the ratio that expresses the relative importance of inertia to the activity that generates spontaneous flows; rather than to viscosity. 
Thus, it is the smallness of $\tilde{\rho}$ that indicates the relevance of inertial non-linearities. 

\section{Characteristic velocity lengths}
\label{app:lengths}

To characterize the various flow states, we examine the spatial velocity-velocity correlation function
\begin{align}
    \label{eq:autoCorr}
    C_{uu}(x) &= \frac{ \av{\vec{u}\left(\vec{r}_0;t\right)\cdot\vec{u}\left(\vec{r}_0+x  \hat{\vec{e}}_\parallel;t\right)} }{ \av{\vec{u}\left(\vec{r}_0;t\right)\cdot\vec{u}\left(\vec{r}_0;t\right)} }
\end{align} 
down the length of the channel for positions $\vec{r}_0$ within $\delta r=\pm2$ of the centre line of the channel. 
The autocorrelations $C_{uu}$ reveal four discernible families of behaviours: \textit{(i)} Those with an initial exponential decay but a constant, positive, long-range correlation; \textit{(ii)} those that pass through $C_{uu}=0$; \textit{(iii)} those with an initial decay but long-range oscillating correlations; and \textit{(iv)} those without long-range oscillating correlations.  

To describe all four behaviours, the autocorrelation functions are fit by the empirical form
\begin{align}
    C_{uu}(x) &= \left(c_1e^{-x/\ell_1}+c_2\right) \cos\left(\frac{2\pi x}{\ell_2}+\phi\right) + c_0. 
    \label{eq:fit}
\end{align} 
The autocorrelation function is a decaying cosine, which so possesses two length scales: a decay length $\ell_1$ and a wavelength $\ell_2$. 
The amplitude of the oscillations decays from $c_1+c_2$ to $c_2$, and the autocorrelation oscillates about $c_0$. 
When the exponential decay does not approach zero, the flow structure has long-range persistence, which is identified in cases for which $c_0$ is significantly greater than zero. 
This absence of any anti-correlation is indicative of a flow state streaming along the channel, which is the case for the axial streaming and grinder states. 
In \fig{fig:corrLength}, cases for which $c_0>1/3$ are represented by open symbols. 
For high activity flow states (Ceilidh lattice and turbulence), $C_{uu}$ crosses zero and $c_0<1/3$, which is denoted by filled symbols. 

In the main text (\fig{fig:corrLength}), the spontaneous flow structures are quantified according to the characteristic length scale $\ell_u$, which is determined by assessing the fit of the correlation functions to \eq{eq:fit}. 
Whether the characteristic length scale $\ell_u$ corresponds to the decay length $\ell_1$ or the wavelength $\ell_2$, is depends on the value of $c_2$. 
We found that all velocity autocorrelations had a short-range drop, which is successfully fit using a decaying cosine of amplitude $c_1$ and decay length $\ell_1$. 
In cases of non-periodic flow (axial flow and turbulence), there is no long-range amplitude and so $c_2\approx0$. 
In these cases, the wavelength is unimportant; thus, the reported length scale is the decay length $\ell_u=\ell_1$ and a circle symbol is used. 
However, both the Ceilidh lattice and grinder train exhibited long-range oscillations in $C_{uu}$ of long-range amplitude $c_2$. 
When  $c_2>0$ the flow states are spatially ordered with translational symmetry breaking (the Ceilidh lattice and grinder train). 
These cases are shown by square symbols in \fig{fig:corrLength}.  
When $c_2>0.05$, the spatial ordering with periodicity characterized by the wavelength $\ell_2$. 
In these cases, the value characteristic lengthscale of the flow is taken to be $\ell_u = \ell_2$. 

\section{Helicity transport equations}
\label{app:helicity}

In order to understand the source of helicity in double helix and grinder train helical structures, we derive the transport equation for helicity for active nematics. 
The local helicity $H=\vec{u}\cdot\vec{\omega}$ is a pseudoscalar field and the integral of $H$ over the entire system is the total helicity, such that $H$ can be referred to as the helicity density field~\cite{moffatt1992}. 
As a first step towards the helicity transport equation, we derive the transport equation for the vorticity. Starting from \eq{eq:mom}, we note that the advection term can be rewritten as $(\vec{u}\cdot\del)\vec{u} = \frac{1}{2}\del u^2 - \vec{u}\times\vec{\omega}$, and take the curl of both sides of the equation. We next employ the fact that $\del\times(\vec{u}\times\vec{\omega}) = \vec{\omega}\cdot\del\vec{u}-\vec{u}\cdot\del\vec{\omega}$ and that the curl of a gradient vanishes, we obtain
\begin{align}
\label{eq:trans_vort}
    \tilde{\rho}D_t\vec{\omega} &= \tilde{\eta}\del^2\vec{\omega}+ \tilde{\rho}(\vec{\omega}\cdot\del)\vec{u} \nn\\
    &\qquad +\del\times\left[\del\cdot(\tilde{K}\tens{\pi}^n-\tens{Q})\right].
\end{align}
This is the non-dimensionalized transport equation for vorticity in an active nematic fluid. 
On the right hand side, the first two terms exist for passive isotropic fluids fluids, whereas the last term is the curl of the divergence of the passive nematic stresses~\cite{Chono1998} and activity.

We can employ the vorticity transport equation to derive the transport equation for the helicity $H = \vec{u}\cdot\vec{\omega}$. 
We consider the rate of change of helicity
\begin{equation}
\label{eq:hel1}
    \partial_t H = (\vec{\omega}\cdot\partial_t\vec{u}+\vec{u}\cdot\partial_t\vec{\omega}), 
\end{equation}
and use \eqs{eq:mom} and \eqref{eq:trans_vort} in \eq{eq:hel1}.
We notice four facts:
\begin{description}
    \item[\textit{(i)}] $\vec{\omega}\cdot(\vec{u}\cdot\del)\vec{u}+\vec{u}\cdot(\vec{u}\cdot\del)\vec{\omega} = (\vec{u}\cdot\del)H$;
    \item[\textit{(ii)}] $\vec{\omega}\cdot\del^2\vec{u}+\vec{u}\cdot\del^2\vec{\omega} = \del^2 H - 2 [\del\vec{\omega}:\del\vec{u}]$;
    \item[\textit{(iii)}] $\del\cdot\vec{\omega} = 0$; and 
    \item[\textit{(iv)}] $\vec{u}\cdot(\vec{\omega}\cdot\del)\vec{u} = \del\cdot\left(\frac{1}{2}u^2\vec{\omega}\right)$. 
\end{description}
Employing these facts, we obtain the dimensionless transport equation
\begin{align}
    \label{eq:transHelicity}
    \tilde{\rho} D_t H +\del \cdot \vec{J} &= \Sigma,
\end{align}
where we have explicitly written the transport equation as resulting from the divergence of flux $\vec{J}$ and source terms $\Sigma$. 
The flux is composed of two terms
\begin{align}
    \label{eq:flux}
    \vec{J} &= -\tilde{\eta} \del H + P'\vec{\omega}, 
\end{align}
where $P'=P-\frac{1}{2}\tilde{\rho}\,u^2$ is an effective pressure.
There are three sources of helicity
\begin{align}
    \label{eq:sources}
    \Sigma &= \tilde{\eta}\Sigma^\text{v} + \tilde{K} \Sigma^\text{n} + \Sigma^\text{a} . 
\end{align}
The first contribution is a viscous source
\begin{align}
    \label{eq:viscSource}
    \Sigma^\text{v} &= - 2 \del \vec{u} : \del \vec{\omega},
\end{align}
which is present in passive isotropic fluids. 
However, the second passive contribution to the helicity is nematohydrodynamic in nature, with the form
\begin{align}
    \label{eq:nemSources}
    \Sigma^\text{n} &= \vec{u}\cdot\left(\del\times\left[\del\cdot\tens{\pi}^\text{n}\right]\right) + \vec{\omega}\cdot\left[\del\cdot\tens{\pi}^\text{n}\right] . 
\end{align}
The Beris-Edwards model of passive nematohydrodynamics is not expected to conserve helicity~\cite{Gay-Balmaz2011}. 
Because $\tilde{K}\sim10^{-3}$ and $\tilde{\eta}\sim10^{-2}$ (\fig{fig:eta}), both the nematic and viscous sources are insignificant. 
Most relevantly, the active contribution to the helicity takes the form
\begin{align}
    \label{eq:actSources}
    \Sigma^\text{a} &= - \vec{u}\cdot\left(\del\times\left[\del\cdot\tens{Q}\right]\right) - \vec{\omega}\cdot\left[\del\cdot\tens{Q}\right] \\
    &= \Sigma^\text{a}_1 + \Sigma^\text{a}_2 \nn. 
\end{align}
The first of these two terms $\Sigma^\text{a}_1=-\vec{u} \cdot (\del\times [\del\cdot\tens{Q}] )$ is the projection of the curl of the active force density ($-\zeta\del\cdot\tens{Q}$) on the velocity. 
On the other hand, the second $\Sigma^\text{a}_2=- \vec{\omega} \cdot [\del\cdot\tens{Q}]$ is the curl of velocity (\ie the vorticity $\vec{\omega}=\del\times\vec{u}$) projected on the force density. 

\begin{figure}[tb]
	\centering
	\includegraphics[width=0.5\textwidth]{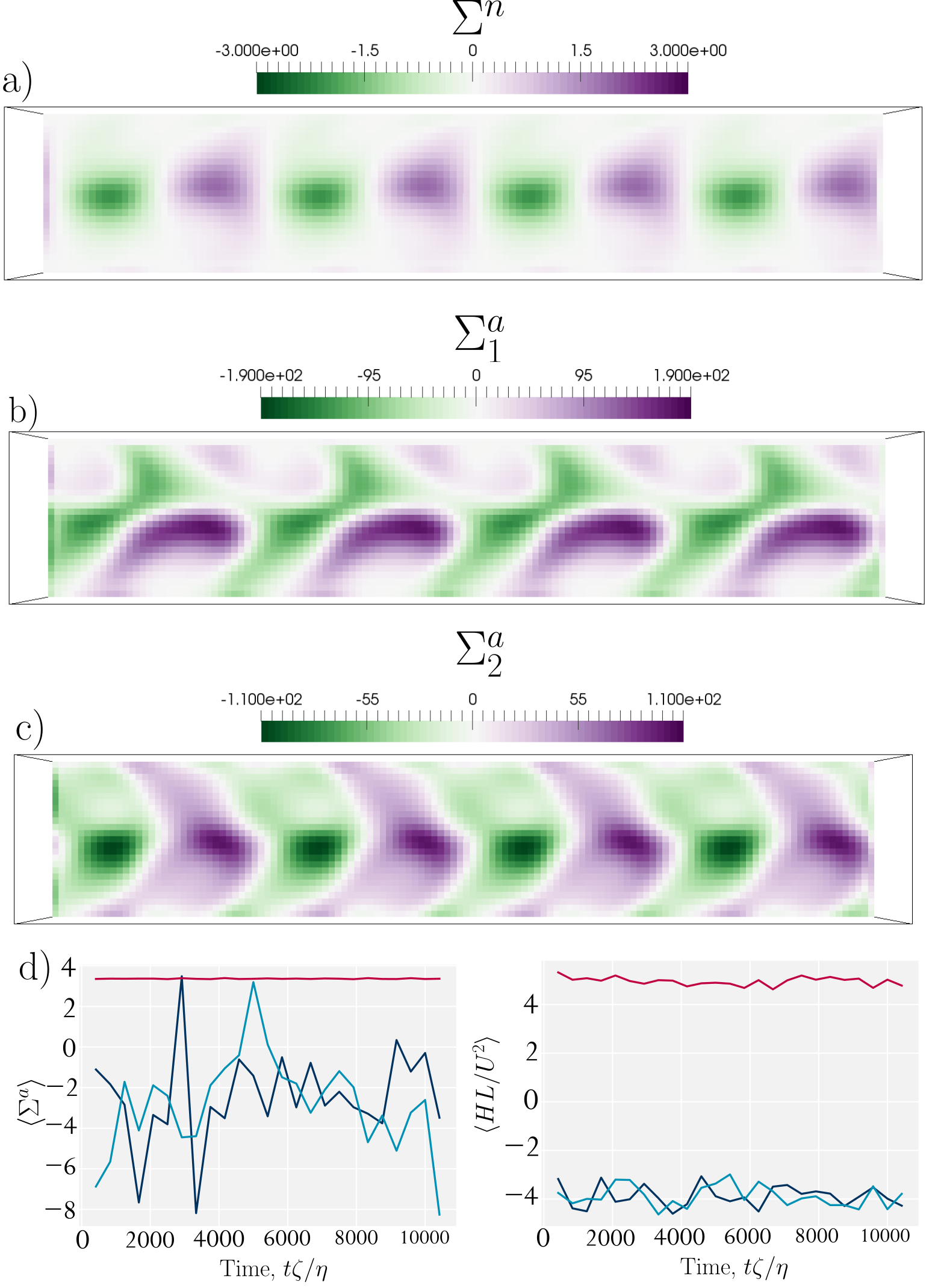}
	\caption{
	 Most significant non-dimensionalized sources of helicity $\Sigma$ for $\tilde{P} = \{ 20,250,0 \}$. 
	(a) Viscous source given by \eq{eq:viscSource}. 
	(b) The $\Sigma^\text{a}_1 = - \vec{u} \cdot (\del\times[\del\cdot\tens{Q}])$ contribution to activity-induced sources (first term from \eq{eq:actSources}). 
	(c) The $\Sigma^\text{a}_2 = - \vec{\omega} \cdot [\del\cdot\tens{Q}]$ contribution to activity-induced sources (second term from \eq{eq:actSources}), the magnitude of which is roughly half the first term. 
	(d) System averaged active helicity source $\av{\Sigma^\text{a}}_\mathbf{r}(t)$ and averaged helicity $\av{HL/U^2}_\mathbf{r}(t)$  as a function of time for three different double helix flow states. The red system occurs at $\tilde{P} = \{ 25, 1.5,9.47 \}$, blue at $\tilde{P} = \{ 25, 0.15,4.7 \}$, and cyan at $\tilde{P} = \{ 25, 0.15,6.5 \}$.
	}
	\label{fig:helicitySources}
\end{figure}

To test the expectation that the non-active contributions to $\Sigma$ are negligible, we measure the viscous source field for the grinder train and compare it to the active terms (\fig{fig:helicitySources}a-c). 
We find that the viscous contribution is two orders of magnitude smaller than the active contribution, which corresponds directly to the scale of $\tilde{\eta}$. 
The viscous source $\tilde{\eta}\Sigma^\text{n} \sim 10^0$ follows the form of the grinder train helicity quite closely (\fig{fig:helicitySources}a). 
The two active sources are seen to be of the same order of magnitude, but the $\Sigma^\text{a}_1 = - \vec{u} \cdot (\del\times[\del\cdot\tens{Q}])$ contribution is roughly twice as strong as the $\Sigma^\text{a}_2 = - \vec{\omega} \cdot [\del\cdot\tens{Q}]$ contribution (\fig{fig:helicitySources}b-c). 
This is found to be true of both the grinder train and double helix.

When the active sources are summed for the grinder train, they are seen to produce a travelling wave of helicity production and elimination. 
However, the net active source of helicity is small and oscillates about zero, explaining why the net helicity in the system is zero. 
On the other hand for the double helix, a net value is found when the active source term is averaged. 
The sign of the average source is found to correspond to the sign of the helicity of each double helix (\fig{fig:helicitySources}d)

\section{Helicity averages}
\label{app:helicityAv}

\begin{figure}[tb]
	\centering
	\includegraphics[width=0.5\textwidth]{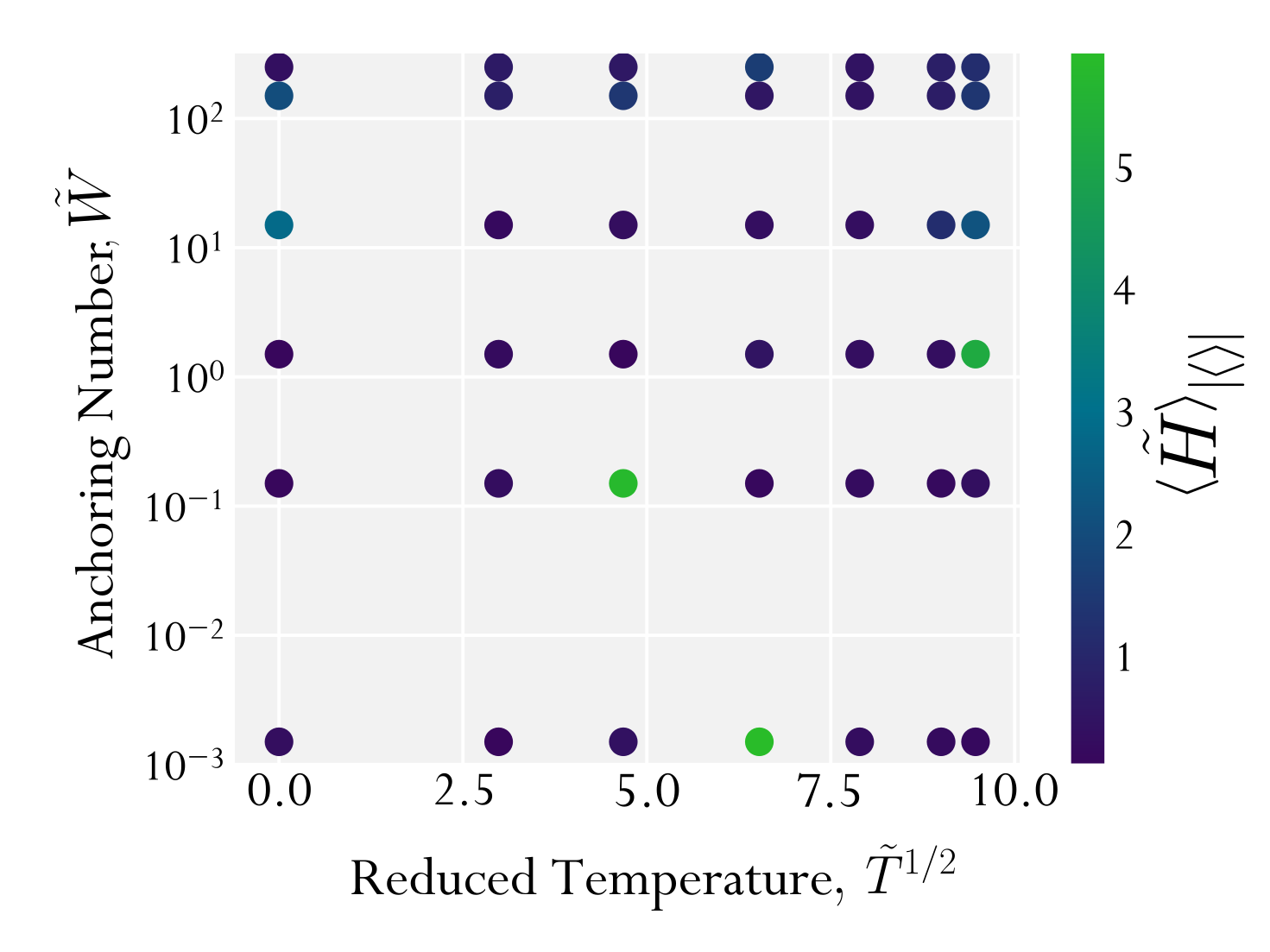}
	\caption{
	Non-dimensionalized, temporal average of the absolute value of the spatially averaged helicity $\langle\tilde{H}\rangle_{|\langle\rangle|}$ given by \eq{eq:absavghelicity}, for activity number $\tilde{K}^{-1/2}=25$. 
	All values are close to zero, except instances of the double helix dissipative structure. 
	}
	\label{fig:absavghelicity}
\end{figure}

In order to differentiate between helical flow states, multiple statistical measures of the helicity field $H\left(\vec{r};t\right)=\vec{u}\cdot\vec{\omega}$ are considered. 
The instantaneous spatial average $\langle H\rangle_\mathbf{r}(t)$ is computed by taking the mean of helicity over space at time $t$. 
To aid comparison in the active dissipative structure phase diagram (\fig{fig:corrLength}c-d), spatially averaged helicty is normalized by $\langle U\rangle^2_\mathbf{r}/L$, where $\langle U \rangle_\mathbf{r}(t)$ is the instantaneous mean fluid speed at time $t$ and $L$ is the confinement length scale. 
Following this normalisation, the absolute value of the non-dimensional temporal average is taken
\begin{align}
    \label{eq:absavghelicity}
    \langle\tilde{H}\rangle_{|\langle\rangle|} &= \left\langle \frac{\abs{\langle H\rangle_\mathbf{r}} }{\langle U\rangle^2_\mathbf{r} / L}\right\rangle_t,
\end{align}
where $\langle\ldots\rangle_t$ denotes the temporal average. 
In \eq{eq:absavghelicity}, the spatial average $\langle H\rangle_\mathbf{r}(t)$ is taken for the entire channel and the temporal average of the absolute value of the spatial average is evaluated.
This is appropriate for the double helix flow states, which spontaneously has positive or negative helicity. However, in the case of the grinder train, the helicity oscillates symmetrically about zero in space (see \fig{fig:helicity}c) and so $\langle\tilde{H}\rangle_{|\langle\rangle|} \simeq 0$.  
Thus, $\langle\tilde{H}\rangle_{|\langle\rangle|} > 0$ highlights the double helix in \fig{fig:absavghelicity}. 
As a cutoff value to obtain the $\otimes$ markers in \fig{fig:corrLength}d, we use $\langle\tilde{H}\rangle_{|\langle\rangle|}>4$. 

Because it is the temporal average of the absolute value of the spatial average, $\langle\tilde{H}\rangle_{|\langle\rangle|}=0$ for the grinder state. 
Thus, we also consider a separate measure of helicity by taking the modulus before spatially averaging
\begin{align}
    \label{eq:avgabshelicity}
    \langle\tilde{H}\rangle_{\langle||\rangle} &= \bigg\langle\frac{\langle \abs{H}\rangle_\mathbf{r} L^2}{\langle U\rangle_\mathbf{r}}\bigg\rangle_t, 
\end{align}
where $\langle H\rangle_\mathbf{r}(t)$ has been taken over the channel centreline.
Both the double helicies and grinder trains exhibit non-zero $\langle\tilde{H}\rangle_{\langle||\rangle}$. 

\begin{figure}[tb]
	\centering
	\includegraphics[width=0.5\textwidth]{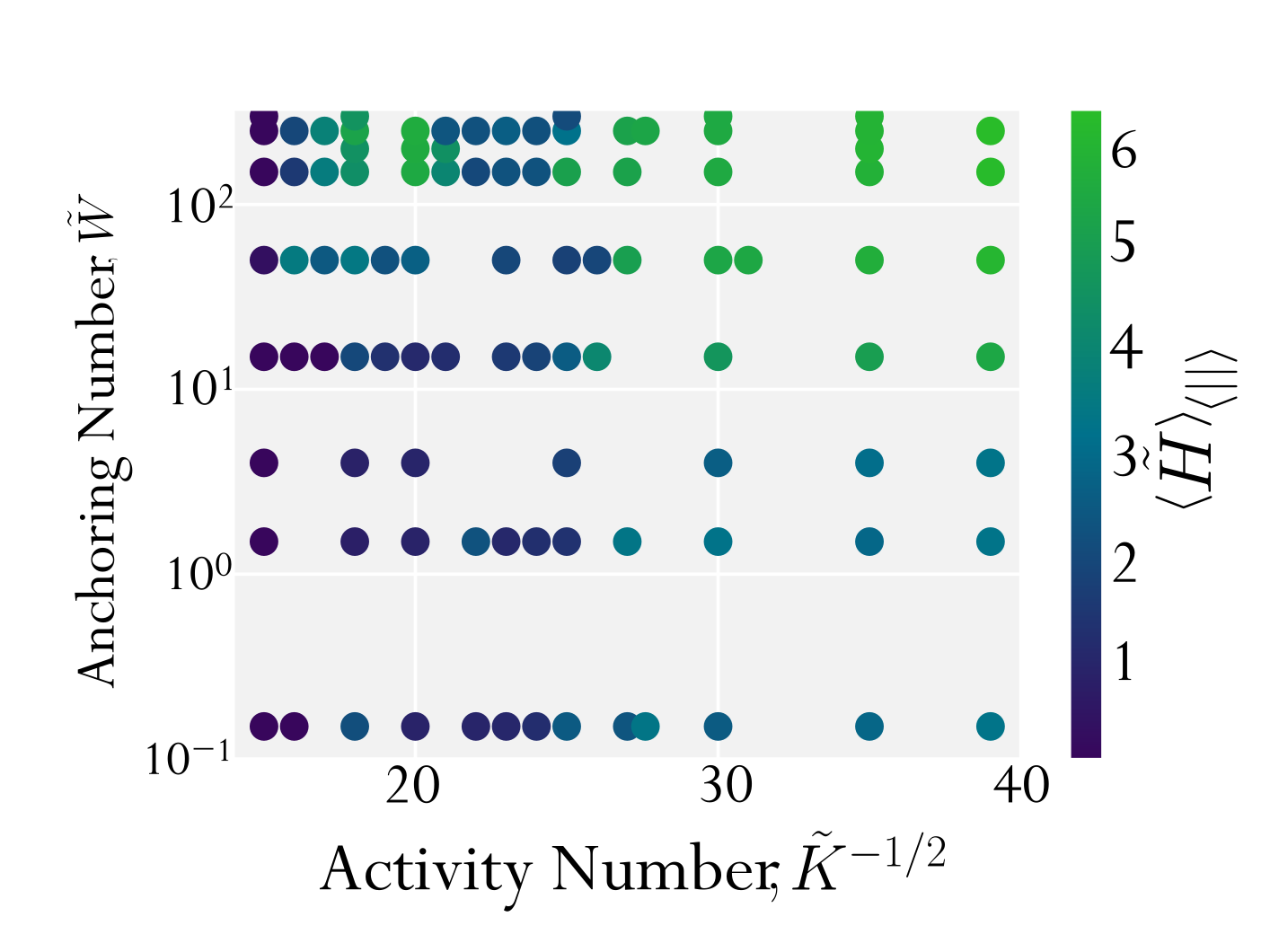}
	\caption{
	Non-dimensionalized, average of the absolute value of helicity $\langle\tilde{H}\rangle_{\langle||\rangle}$  given by \eq{eq:avgabshelicity},  for reduced temperature $\tilde{T}^{1/2}=0$. 
	In addition to non-zero values in regions of active turbulence, the grinder train regime exists as well defined region of large $\langle\tilde{H}\rangle_{\langle||\rangle}$ surrounded by small values at strong anchoring and moderate activity. 
	}
	\label{fig:avgabshelicity_act}
\end{figure}

The grinder train points can be seen in \fig{fig:avgabshelicity_act} as the points at high anchoring and moderate activity $\tilde{K}^{-1/2} \approx 20$ with $\langle\tilde{H}\rangle_{|\langle\rangle|}>4$. 
The average helicity $\langle\tilde{H}\rangle_{\langle||\rangle}$ is small in the surrounding parameter space. 
However, in addition to the grinder train, there exists an extended region of large $\langle\tilde{H}\rangle_{\langle||\rangle}$ at high activity and high anchoring number --- this is the region of active turbulence. 
The turbulent regime exhibits large values of $\langle\tilde{H}\rangle_{\langle||\rangle}$ owing to high fluid speeds. 
To differentiate between the grinder train and turbulent flow states, we would thus need to combine the helicity measure with the periodic nature of the velocity autocorrelation functions (\eq{eq:autoCorr}). 
However, we find that peak resolution (\appndx{app:peakRes}; \eq{eq:res}) is a clearer metric. 

\begin{figure}[tb]
	\centering
	\includegraphics[width=0.5\textwidth]{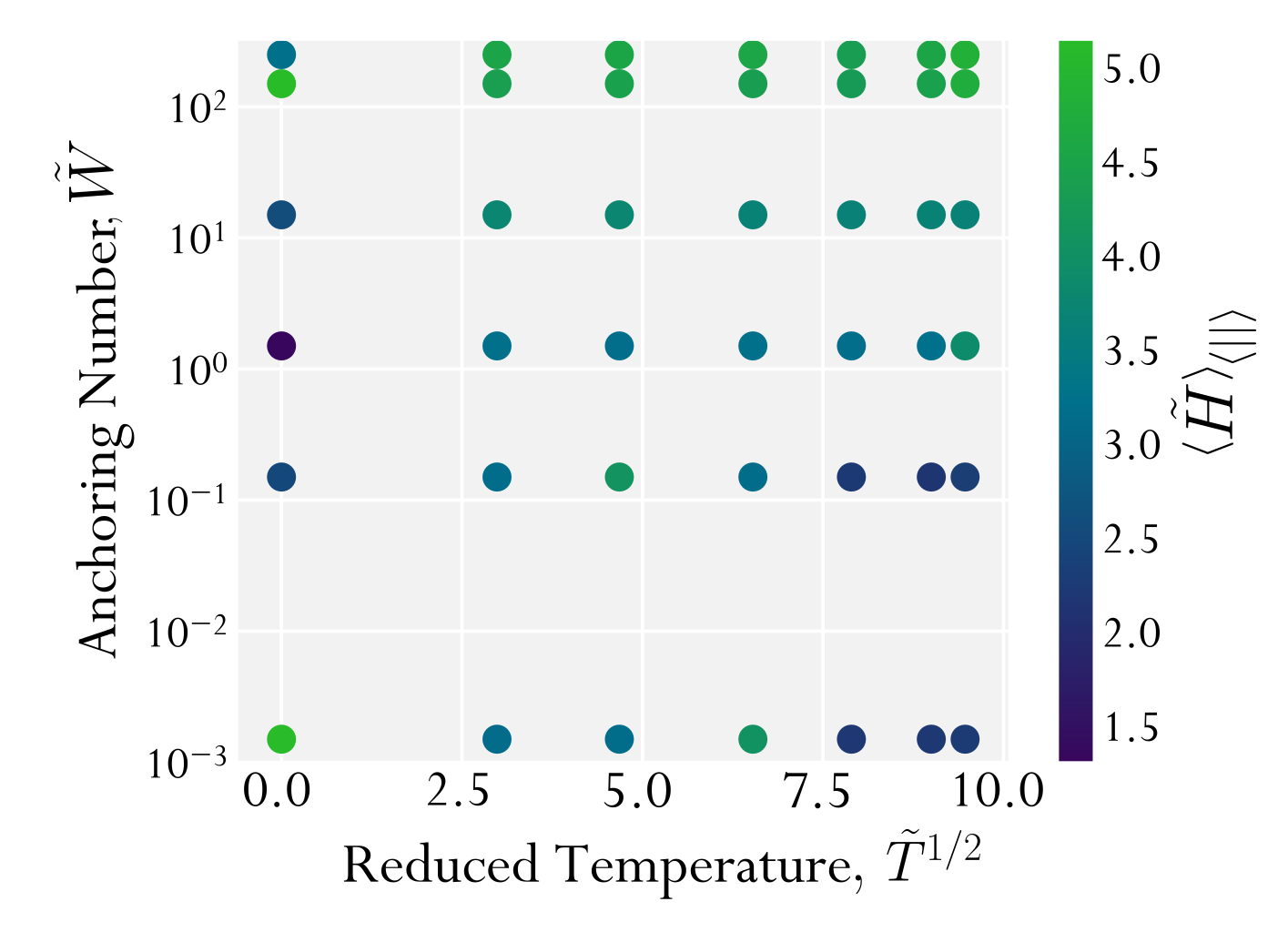}
	\caption{
	Non-dimensionalized, average of the absolute value of helicity $\langle\tilde{H}\rangle_{\langle||\rangle}$  given by \eq{eq:avgabshelicity}, for $\tilde{K}^{-1/2}=25$. 
	Non-zero values exist for the double helix, but also active turbulent regions of parameter space. 
	}
	\label{fig:avgabshelicity_temp}
\end{figure}

For the double helix both $\langle\tilde{H}\rangle_{|\langle\rangle|}$ (\fig{fig:absavghelicity}) and $\langle\tilde{H}\rangle_{\langle||\rangle}$ (\fig{fig:avgabshelicity_temp}) are large. 
However, the intermittent double helix, which exists for high reduced temperatures and anchoring numbers, has near zero $\langle\tilde{H}\rangle_{|\langle\rangle|}$. 
As cutoffs for the intermittent double helix, marked by $\oslash$ in \fig{fig:corrLength}d, we check that $\langle\tilde{H}\rangle_{|\langle\rangle|}<1.0$ but $\langle\tilde{H}\rangle_{\langle||\rangle}>4.5$, which is a value that is even greater than in the orderly double helix. 

\section{Helicity Peak Resolution}
\label{app:peakRes}

\begin{figure}[tb]
	\centering
	\includegraphics[width=0.5\textwidth]{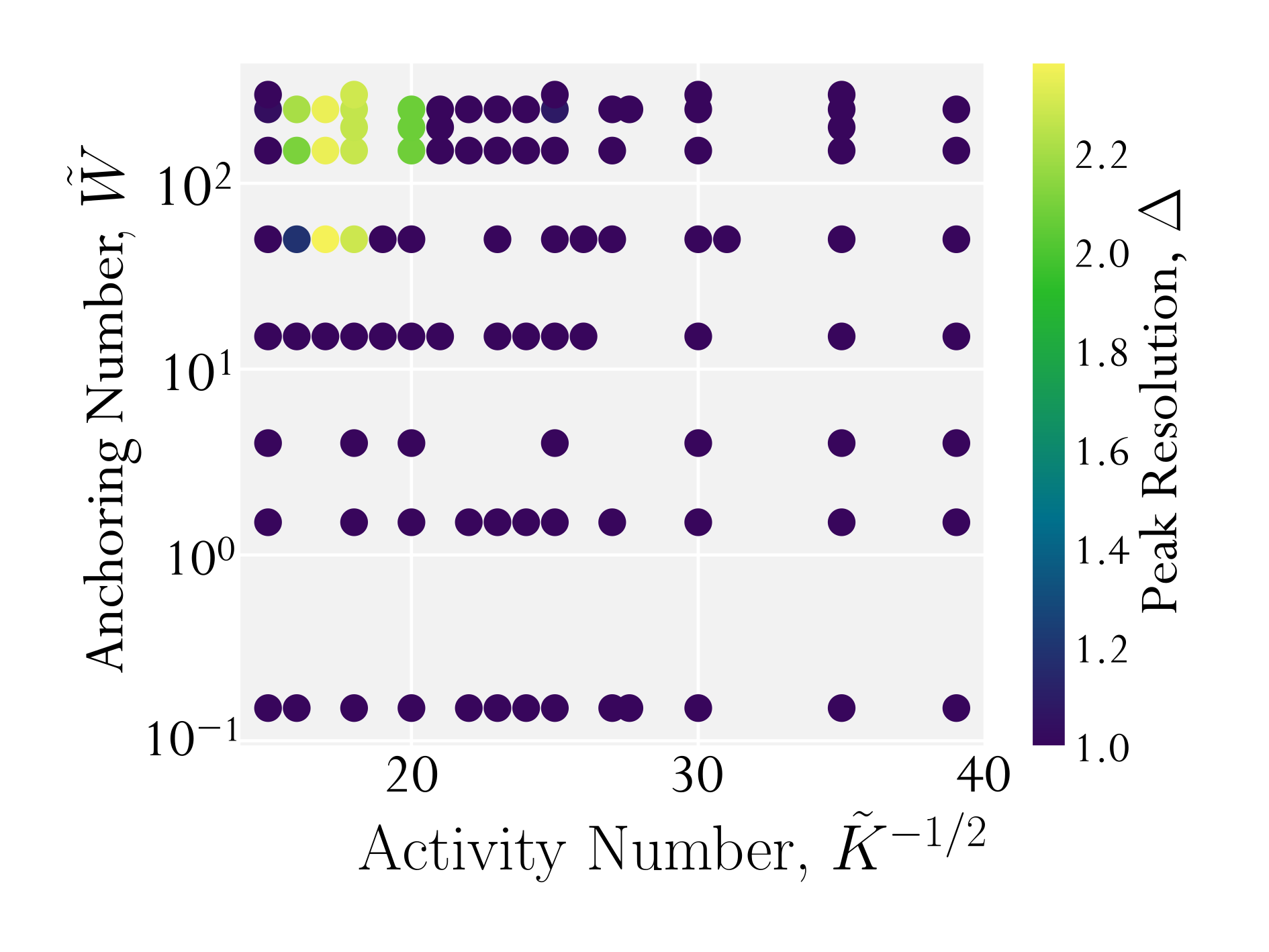}
	\caption{
	Peak resolution $\Delta$ of the helicity distributions given by \eq{eq:res},  for reduced temperature $\tilde{T}^{1/2}=0$. 
	Only the grinder train regime possesses significant bimodality to produce large peak resolution. 
	}
	\label{fig:res_act}
\end{figure}

The distributions of helicity across the examined parameter space are generally Gaussians, symmetric about zero. 
However, this is not true of the helical regimes presented in \fig{fig:helicity}b of the main text. 
The double helix does have a Gaussian distribution of helicity but with non-zero mean. 
Furthermore, the grinder train possesses a strongly peaked, symmetric, bimodal distribution of helicity, which can be fit to the sum of two skewed normal distributions. To each of the peaks in the distribution we find a fit of the form 
\begin{align}
    \mathcal{G}_i &= A_i \exp\left(-\frac{(x-\mu_i)^2}{2\sigma_i^2}\right)\left[1+\textrm{erf}\left(\frac{S^k_i(x-\mu_i)}{\sigma_i\sqrt{2}}\right)\right]. 
    \label{eq:dist}
\end{align}
The peaks are defined by their amplitude $A_i$, mean $\mu_i$, standard deviation $\sigma_i$, and skew $S^k_i$, $i=1,2$. 
To quantify if the two peaks are separated, and thus the distribution bimodal, we define the peak resolution as
\begin{align}
    \Delta &= \frac{\abs{\mu_1 - \mu_2}}{2\sqrt{\sigma_1\sigma_2}} . 
    \label{eq:res}
\end{align} 
It is found that only the grinder train flow states possess strong bimodal helicity (\fig{fig:res_act}). 
We mark the grinder train with $\oplus$ in \fig{fig:corrLength}c when $\Delta>1.2$. 

\section{Movie Captions}
\label{app:movies}
\begin{enumerate}
    \item\label{mov:DH} Double helix flow state for parameters $\tilde{P} = \{ \tilde{K}^{-1/2},\tilde{W},\tilde{T}^{1/2} \} = \{ 25,1.5,9.47\}$. Two streams of anti-parallel flow braid around one another. 
    Coloured by axial velocity $u_\parallel$. Initialized at rest and isotropic. 
    \item\label{mov:GT} Grinder train for $\tilde{P} = \{ \tilde{K}^{-1/2},\tilde{W},\tilde{T}^{1/2} \} = \{ 20,250,0 \}$. A Procession of spontaneously drifting, counter-rotating vortices aligned axially down the duct. Coloured by axial vorticity $\omega_\parallel$. Initialized at rest and isotropic. 
 	\item\label{mov:VL} Ceilidh vortex for $\tilde{P} = \{ \tilde{K}^{-1/2},\tilde{W},\tilde{T}^{1/2} \} = \{ 22,250,0 \}$. Vortex lattice oriented in the transverse direction do not exhibit net axial flow. Coloured by transverse vorticity $\omega_\perp$. Initialized at rest and isotropic. 
    \item\label{mov:disc} Disclinations for the flow states shown in \fig{fig:flows} and \movie{mov:DH}-\ref{mov:VL}). 
    Colourmaps show the same quantities as in \fig{fig:flows} in the central plane. 
    \textbf{Top:} The double helix flow state is accompanied by a single helical disclination line that turns like a corkscrew. 
    \textbf{Middle:} The grinder train is entirely devoid of defects. 
    \textbf{Bottom:} The vortex lattice possesses Ceilidh dancing defects which in 3D are arced disclination lines, as well as stationary disclination lines at the walls. Time is stated in units of $10^2$ LB time steps.
    \item\label{mov:h-GT}  Local helicty density field for the grinder train for $\tilde{P} = \{ \tilde{K}^{-1/2},\tilde{W},\tilde{T}^{1/2} \} = \{ 20,250,0 \}$. Initialized at rest and isotropic. 
    \item\label{mov:h-DH} Local helicty density field for the double helix flow state for parameters $\tilde{P} = \{ \tilde{K}^{-1/2},\tilde{W},\tilde{T}^{1/2} \} = \{ 25,0.0015,6.5\}$. Initialized at rest and isotropic. 
    \item\label{mov:IDH} Intermittent double helix flow state for parameters $\tilde{P} = \{ \tilde{K}^{-1/2},\tilde{W},\tilde{T}^{1/2} \} = \{25,250,9.47\}$. More chaotic behaviours, in which one helix sporadically dominates or the helical structure temporarily disperses. Coloured by axial velocity $u_\parallel$. Initialized at rest and isotropic. Time intervals in units of $10^3$ LB time steps between frames. 
    \item\label{mov:GT-actSource} Active source of helicity $\Sigma^\text{a}$ for the grinder train (parameters $\tilde{P} = \{ \tilde{K}^{-1/2},\tilde{W},\tilde{T}^{1/2} \} = \{20,250,0\}$). 
    Red and blue contours show $\Sigma^\text{a}=\pm175$. 
    Time is stated in units of $10^2$ LB time steps. 
    Initialized at rest and isotropic.
    \item\label{mov:DH-actSource} Active source of helicity $\Sigma^\text{a}$ for the double helix (parameters $\tilde{P} = \{ \tilde{K}^{-1/2},\tilde{W},\tilde{T}^{1/2} \} = \{25,1.5,9.47\}$. 
    Red and blue contours show $\Sigma^\text{a}=\pm80$. 
    The disclination lines are highlighted as yellow contours of $S=0.2$. 
    Time is stated in units of $10^2$ LB time steps. 
    Initialized at rest and isotropic. 
\end{enumerate}

\bibliography{refHelix}

\end{document}